\documentclass[prl,twocolumn,showpacs,floatfix,superscriptaddress]{revtex4-1}
\usepackage{graphicx,color}

\begin{document}

\title{Electronic and structural characterization of divacancies in irradiated graphene }ºº

\author{Miguel M. Ugeda}
\email[Corresponding author: ]{miguel.moreno@uam.es}
\affiliation{Dept.\ F\'{\i}sica de la Materia Condensada, Universidad Aut\'onoma de Madrid, E-28049 Madrid, Spain}

\author{Iv\'an Brihuega}
\affiliation{Dept.\ F\'{\i}sica de la Materia Condensada, Universidad Aut\'onoma de Madrid, E-28049 Madrid, Spain}

\author{Fanny Hiebel}
\affiliation{Institut N\'eel, CNRS-UJF, BP166, 38042 Grenoble, France}

\author{Pierre Mallet}
\affiliation{Institut N\'eel, CNRS-UJF, BP166, 38042 Grenoble, France}

\author{Jean-Yves Veuillen}
\affiliation{Institut N\'eel, CNRS-UJF, BP166, 38042 Grenoble, France}

\author{Jos\'{e} M.\ G\'omez-Rodr\'{\i}guez}
\affiliation{Dept.\ F\'{\i}sica de la Materia Condensada, Universidad Aut\'onoma de Madrid, E-28049 Madrid, Spain}

\author{F\'elix Yndur\'ain}
\affiliation{Dept.\ F\'{\i}sica de la Materia Condensada, Universidad Aut\'onoma de Madrid, E-28049 Madrid, Spain}

\begin{abstract}
We provide a thorough study of a carbon divacancy, a fundamental but almost unexplored point defect in graphene. Low temperature scanning tunneling microscopy (STM) imaging of irradiated graphene on different substrates enabled us to identify a common two-fold symmetry point defect. Our first principles calculations reveal that the structure of this type of defect accommodates two adjacent missing atoms in a rearranged atomic network formed by two pentagons and one octagon, with no dangling bonds. Scanning tunneling spectroscopy (STS) measurements on divacancies generated in nearly ideal graphene show an electronic spectrum dominated by an empty-states resonance, which is ascribed to a spin-degenerated nearly flat band of $\pi$-electron nature. While the calculated electronic structure rules out the formation of a magnetic moment around the divacancy, the generation of an electronic resonance near the Fermi level, reveals divacancies as key point defects for tuning electron transport properties in graphene systems.
\end{abstract}


\maketitle

Graphene is a unique material from an electronic point of view. The ultrarelativistic nature of its charge carriers \cite{Novoselov2005} and the robustness of its electronic coherence \cite{Novoselov2004} make it the ideal candidate in the forthcoming era for nanoelectronics. However, all these remarkable electronic and transport properties are subjected to the presence of ubiquitous disorder. By instance, extrinsic/intrinsic defects commonly present in graphene are considered the limiting factor for electronic transport through charged impurities \cite{Adam2003,Tan2007}, rippling \cite{Katsnelson2008} or resonant scatterers \cite {Titov2010,Ni2010}. These defects lead to substantial changes in the topology of its low-energy electronic bands, which are at the origin of many of its unique properties. Taking advantage of the key role of defects in low dimensional carbon systems, a new route based on defect-engineering is being developed to broaden the functionalities of graphene \cite{Lusk2008,Banhart2011}. Adatoms, vacancies and Stone-Wales (SW) defects are the most common defects invoked to act as building blocks in structurally tailored graphene. Both vacancy-type and SW defects can be artificially created in graphene by electron or ion irradiation and visualized with atomic resolution by high-resolution transmission electron microscopy \cite{Meyer2008} and scanning probe microscopy \cite{Kelly1996,Ruffieux2000}. In particular, STM has demonstrated to be a powerful tool to explore the local electronic structure of such point defects at the atomic scale \cite{Ugeda2010,Ugeda2011,Cockayne2011}.

The potential effects of topological defects in carbon-based systems have been comprehensively studied by theory in the last years, predicting already new functional properties like the opening of an electronic gap in the Dirac bands in graphene with SW defects \cite{Peng2008} or aiming to explain the unexpected magnetic ordering \cite{Esquinazi2003,Ramos2010} observed in defective carbon systems \cite{Lehtinen2004,Vozmediano2005,Pereira2006,Amara2007,Palacios2008,Yazyev2008,Zhang2011}. Therefore, a detailed picture of all types of point defects existing in graphene is mandatory in order to understand their particular impact on the material’s properties and to be able to selectively tailor them. From the experimental point of view, although STM is an ideal technique to study the properties of point defects in graphene systems as single entities, few works are yet reported, mainly restricted to single atomic vacancies. 

In this Letter, we present a combined experimental and theoretical characterization of a divacancy, a point defect expected to be stable and common in ion irradiated carbon samples \cite{Krasheninnikov2001,Lehtinen2010}. Divacancies are also expected to be of fundamental relevance regarding the electron transport properties of these systems \cite{Lherbier2011} being considered, for instance, to be the key defect in order to tune the conductance of carbon nanotubes \cite{Gomez-Navarro2005}.The experimental data here reported were acquired at 6K by using a home-made low temperature STM (LT-STM) in ultra-high-vacuum (UHV) conditions \cite{Ugeda2011b}. The local density of states (LDOS) of the sample was studied with atomic precision by measuring differential conductance (dI/dV) spectra using the lock-in technique (f = 2.3 kHz, $V_{mod}$ = 1.5 mV). All STM/STS data were acquired and processed using the WSxM software \cite{Horcas2007}. In this work we have performed various density functional \cite{Kohn1965} calculations (DFT) using the SIESTA code \cite{Ordejon1996,Soler2002} which uses localized orbitals as basis functions \cite{Sankey1989}. We use a double $\zeta$ polarized basis set, non-local norm conserving pseudopotentials and a functional for the exchange and correlation including van der Waals interactions \cite{Dion2004,Kong2009}. The calculations are performed with stringent criteria in the electronic structure convergence (down to $10^{-5}$ in the density matrix), Brillouin zone sampling (up to 900 k-points), real space grid (energy cut-off of 500 Ryd) and equilibrium geometry (residual forces lower than 0.02 eV/\AA). 

\begin{figure}
\includegraphics[width=85mm]{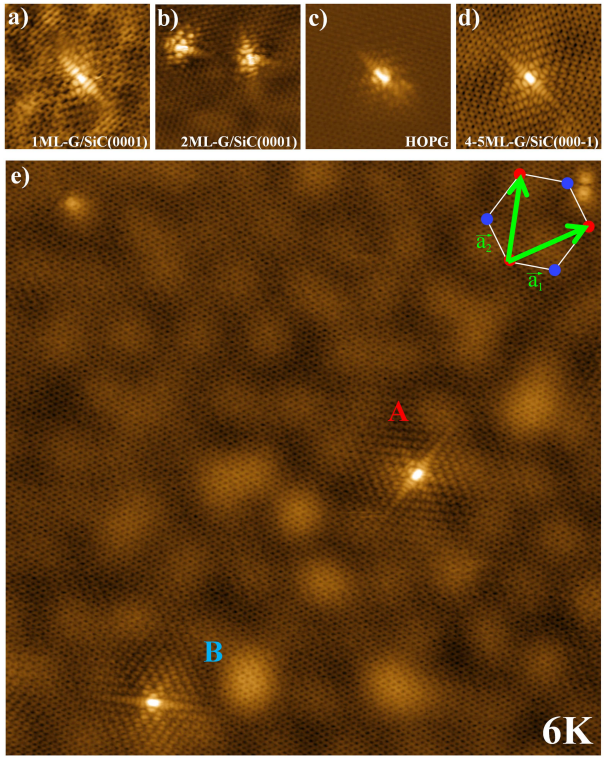}
\caption{\label{Fig1} (color online)  Divacancy in four different graphene systems: (a) 1ML-G/SiC(0001) (V=+20 mV), (b) 2ML-G/SiC(0001), (V=+500 mV), (c) HOPG (V=+500 mV) and (d) 4-5ML-G/SiC(000$\overline{1}$) ( V=+190 mV) . (a)-(d) Size = 6x6 nm${^2}$. All bias voltages refer to the sample. (e) STM image (26 x 26 nm${^2}$) showing two divacancies in 4-5ML-G/SiC(000$\overline{1}$) (V=+280 mV). The direction of graphene lattice vectors $\bf{a_{i}}$ of the outermost layer is indicated at the top right part of the image. }
\end{figure}

We deliberately generated divacancies in several graphene systems by means of Ar${^+}$  irradiation in UHV (140eV), following the same experimental procedure as in references \cite{Ugeda2010,Ugeda2011}. We chose four different graphene systems: monolayer and bilayer graphene on SiC(0001), 4-5 layers of graphene grown on SiC(000$\overline{1}$) and the surface of UHV-exfoliated highly oriented pyrolitic graphite (HOPG). The methods used to grow epitaxial graphene on both SiC orientations (Si and C face) are described in \cite{Mallet2007,Brihuega2008} and \cite{Varchon2008} respectively. After the irradiation procedure we imaged the samples by LT-STM to explore the created defects. Among the artificially generated defects, we were able to identify a common point structure existing in all these graphene surfaces, which is displayed in Figs. 1(a)-(d). According to the STM images, the most characteristic feature of such defect is a two-fold symmetry involving a complex electronic pattern. Another common characteristic of all these defects is the orientation of the mirror plane parallel to their long axis which is rotated $30\,^{\circ}{\rm}$ with respect to the graphene lattice vectors (see fig 1(e)). The similar shape of the defects and their equal orientation with respect to the honeycomb lattice, strongly suggest that they all belong to the same kind of vacancy-type defect. Interestingly, the occurrence of this type of defect in each one of the graphene systems studied here turns out to be quite different. While for the SiC systems this kind of defect was frequently observed after the irradiation procedure, in the case of HOPG surfaces such defect was extremely rare, indeed, the defect shown in Fig. 1(c) was the only one found on the HOPG surface after extensive measurements in many different irradiated HOPG surfaces. The reasons explaining this strong unbalance in the divacancy population remain unclear to us and they are most likely related to the different underlying environment for each graphene system.

In order to further investigate this kind of defect we focused on graphene grown on the C terminated face of SiC, where the rotational disorder of the graphene layers electronically decouple π bands for a large interval of rotational angles \cite{Hass2008,Li2010,x}. This can lead to a stacking of undistorted and nearly isolated graphene sheets in this system, as shown by the experimental observation of ideal non-doped graphene cones around K points \cite{Sprinkle2009}. Therefore, surface STM measurements on multilayer graphene on SiC(000$\overline{1}$) can be considered as a model experimental approximation to the study of ideal neutral monolayer graphene. Figure 1(e) shows a large scale STM image of an irradiated surface of 4-5ML-graphene/SiC(000$\overline{1}$) with two of such vacancies present in the terrace (labeled as A and B). In addition to the atomic graphene lattice, two other super-periodicities can be observed in the STM topograph. Such modulations, known as moir\'e patterns \cite{Pong2005}, are frequently found in graphene grown on SiC(000$\overline{1}$) surfaces \cite{Varchon2008} and they are due to the rotational misalignment of the upmost graphene layers. The moir\'es observed here arise from two different rotations of $20\,^{\circ}{\rm}$ and $5\,^{\circ}{\rm}$ between graphene layers from an AA stacking leading to a periodicity of 7.1\AA and 29.5\AA,  respectively. The larger moir\'e is compatible with a commensurate superstructure (n,m) = (6,7) with a moir\'e vector $\bf{V}$= n$\bf{a_{1}}$  m$\bf{a_{2}}$, $\bf{a_{i}}$ being the graphene lattice vectors (see inset fig 1e) and n,m integers. 

\begin{figure*}
\includegraphics{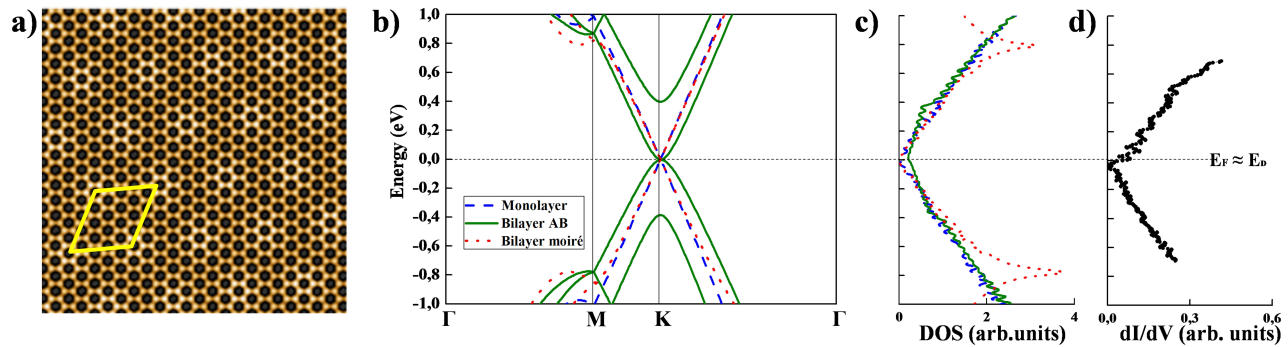}
\caption{\label{Fig2} (color online) (a) Simulated STM image of a commensurate moir\'e (2,3) formed by two graphene layers rotated $\Theta$ = $13.17\,^{\circ}{\rm}$ from an AA stacking. The unit cell (yellow) has a periodicity of 10.7\AA. Parameters: 5x5nm${^2}$, V = +200mV. (b) Electronic band structure calculated for monolayer graphene (dash blue line), bilayer AB (green line) and the (2,3) moir\'e (red dot line). (c) Corresponding DOS for the calculated structures. d) Typical dI/dV spectrum acquired on the surface of 4-5ML-Graphene/SiC(000$\overline{1}$) at 6K. Plots showed in (c) and (d) have the same energy range than in figure (b). }
\end{figure*}

The electronic decoupling due to the rotation between graphene layers is illustrated in figure 2. First, we have calculated a bilayer graphene structure forming a (2,3) commensurate moir\'e using the SIESTA code, as indicated above. Figure 2(a) shows a simulated STM image, within the Tersoff-Hamann approximation \cite{Tersoff1983}, of the relaxed structure, in which the moir\'e can be readily observed. The calculated moir\'e corrugation (5.3 pm) results from both topographical (3.2 pm) and electronic (2.1 pm) contributions which are in phase. The corresponding low-energy electronic band structure of this moir\'e bilayer is represented in fig. 2(b), together with the calculated ones for the monolayer and AB bilayer. While the linear π band dispersion of monolayer graphene becomes quadratic in the vicinity of the Dirac point ($E_{D}$) in a Bernal AB bilayer, for such energies the electronic structure of graphene is restored in the bilayer when both layers are rotated, as already reported \cite{Hass2008,Li2010,x}. Interestingly, although the calculated DOS shows in all cases a similar "V" shape in this energy range (see fig. 2(c)), only the rotated bilayer shares a vanishing DOS at $E_{D}$ with the monolayer graphene. Our tunneling spectroscopy experiments performed on different moir\'es in graphene on SiC(000$\overline{1}$) agree with these calculations. Figure 2(d) shows a dI/dV curve measured on a pristine region. The typical V-shaped and the vanishing differential conductance at $E_{D}$ (in this case at -20mV with respect to the Fermi energy ($E_{F}$) due to residual n-doping of our SiC(000$\overline{1}$) samples) confirms an effective decoupling of the graphene surface layer with respect to the substrate, allowing a direct comparison of the experimental data with calculated structures for a single monolayer.

\begin{figure}
\includegraphics{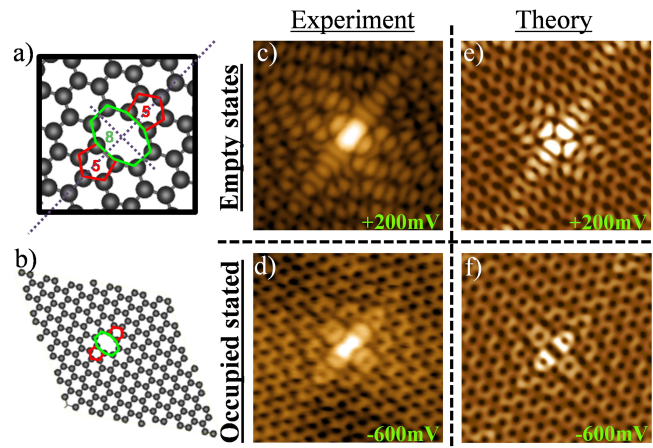}
\caption{\label{Fig3} (color online) (a) Relaxed structure of the calculated divacancy where all sigma bonds saturate leading to the (585) structure. Dashed lines show the mirror planes in the defect. (b) 2.57 x 2.57 nm${^2}$ unit cell used for the calculation. The right panel shows the comparison between experimental [(c) and (d)] and calculated [(e) and (f)] STM images at both sample bias polarities. All images have a size of 3.3 x 3.3 nm${^2}$. }
\end{figure}

The experimental results presented in figure 1 do not reveal unambiguously the atomic structure of the studied defect. The main contribution to the STM images around defects is electronic so they only provide indirect fingerprints of the atomic rearrangement. However, the structure of the defect likely involves vacancy sites as ion irradiation on graphene systems mainly produces voids in the atomic network. More precisely, recent ion bombardment simulations on graphene performed by Lehtinen \emph{et al.}, predict that low-energetic impacts should mostly produce single and double vacancies rather than more complex structures like Frenkel pairs or SW defects.\cite{Lehtinen2010}. The calculated ratio of mono- and divacancies in graphene, according to our irradiation parameters (E(Ar${^+}$) = 140 eV at  $20\,^{\circ}{\rm}$ off normal), is roughly 2:1 \cite{Lehtinen2011}. As stated previously, single atomic vacancies are the only well studied point defects in graphene systems by STM, showing a characteristic electronic pattern with a well defined three-fold symmetry \cite{Kelly1996,Ruffieux2000,Ugeda2010}, in contrast to the present defect. Therefore, we performed DFT calculations focused on divacancies structures to complete the characterization of the two-fold symmetry vacancy-type defect here observed. There are various possible atomic configurations saturating all dangling bonds from a honeycomb lattice where two carbon atoms are removed. The resulting lattices are locally formed by non hexagonal polygons that saturate all $\sigma$ dangling bonds. The simplest of these atomic structures for a reconstructed divacancy is the (585) defect, shown in figure 3(a). The rearranged planar structure encloses a central octagon and two opposing fivefold rings. We have chosen this configuration as the initial structure for the calculation due to its two-fold symmetry and the relative orientation of the two orthogonal mirror planes to the honeycomb lattice, $0\,^{\circ}{\rm}$ and $30\,^{\circ}{\rm}$, as measured in the STM images. Indeed, previous DFT calculations of this divacancy in the LDA approximation display similarities with the present experimental images \cite{Amara2007}. In our monolayer supercell DFT calculation, and in order to minimize the interaction between defects in neighboring cells, we have considered large skewed unit cells. In particular, we present the results for the unit cell defined by the basis vectors $\bf{b_{1}}$= 5$\bf{a_{1}}$ and $\bf{b_{2}}$= 7$\bf{a_{2}}$ (see Fig. 3(b)). Similar results have been obtained with other choices of the unit cell provided they are large enough. As can be noticed, the simulated images obtained from the relaxed structure (Figs. 3(e)-(f)) capture the main features of the experimental STM images (Figs. 3(c)-(d)) at both polarities: a central bright lobe surrounded by a complex electronic ($\sqrt{3}$x$\sqrt{3}$)R$30\,^{\circ}{\rm}$ pattern with twofold symmetry in an equal orientation. This excellent agreement allows us to unambiguously identify this defect with a (585) divacancy. 

\begin{figure}
\includegraphics{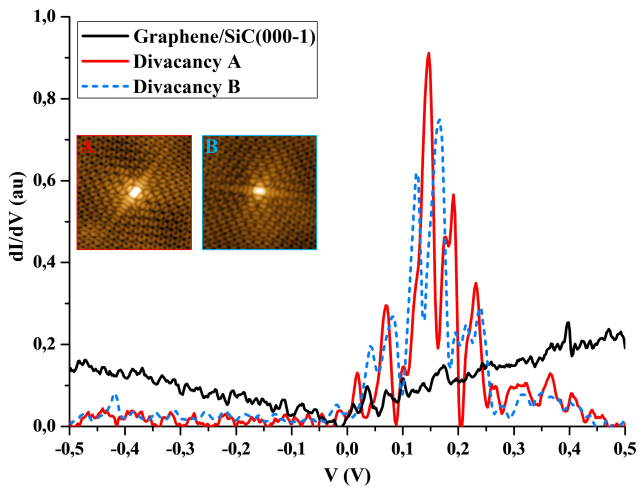}
\caption{\label{Fig4} (color online) Consecutive tunneling differential conductance spectra acquired on the divacancies labeled as A and B in fig. 1(e). A reference dI/dV curve taken on the bare surface is also shown. }
\end{figure}

Topological defects in graphene systems break the translational symmetry of the crystal leading to profound changes in the low-energy electronic bands. Thus, to complete the characterization of this class of divacancies, we have investigated their local electronic structure by means of STS experiments and DFT calculations. Recent theoretical works \cite{Vozmediano2005,Pereira2006,Palacios2008,Yazyev2008} have predicted the existence of quasilocalized states in the vicinity of the $E_{F}$ for different types of topological defects in graphene such atomic vacancies, point impurities or edges. Indeed, these electronic modifications are expected to be at the heart of the striking behavior observed in functionalized graphene regarding its transport and magnetic properties. These states have been experimentally observed by some of us in single atomic vacancies in HOPG and graphene on Pt(111) surfaces \cite{Ugeda2010,Ugeda2011}. Our STS findings in the present case are summarized by the local tunneling spectra taken at the center of the A and B divacancies on the SiC(000$\overline{1}$) region shown in fig. 1(e). Figure 4 shows dI/dV curves acquired consecutively on both A (continuous red) and B (point blue) defects within 0.5 eV around $E_{F}$ together with a reference curve measured with the same tip on a neighboring pristine graphene region. While the latter spectrum shows a "V" shape with a DOS vanishing at $E_{D}$ $\approx$ $E_{F}$ characteristic of ideal graphene as discussed above, it is clear that the presence of vacancies profoundly alter the LDOS. dI/dV spectra recorded on A and B divacancies show a resonance located in empty states and centered at +150mV, which present a rich internal structure. Although both defects share the main features in their electronic spectra, and thus the LDOS reflects qualitatively an equal electronic structure, some subtle differences in the internal structure of the resonances were systematically distinguished. The calculated band structure for a divacancy in a monolayer with a (5,7) unit cell is shown in Fig. 5(a) along with the band structure of the defect free monolayer. In this figure the effect of the presence of the divacancy is apparent; it induces an almost dispersionless $\pi$-character band interacting with the continuum, giving rise to a 0.3 eV width asymmetric resonance in the DOS (see Fig. 5(a)). Thus, as can be noted in Figure 5(b), the calculated total DOS for the (585) divacancy is in nice agreement with the experimental dI/dV curves. However, although the width of the resonances are similar in both cases, the internal structure of the experimental resonance is not clearly reflected in the calculated DOS. Highly demanding further calculations including a second rotated graphene layer and involving a very large number of carbon atoms could be required in order to understand this subtle effect. Finally, although we have performed a spin-resolved calculation, the obtained charge distribution is identical for both spin states, leading therefore to a non magnetic solution.

\begin{figure}
\includegraphics{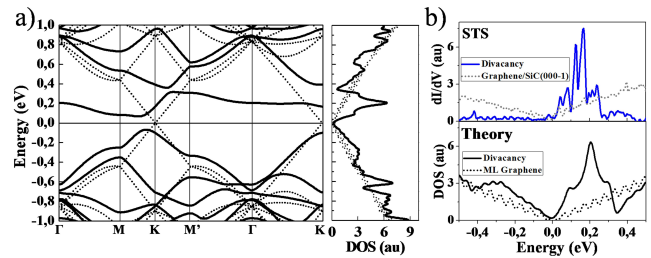}
\caption{\label{Fig5} (color online) (a) Calculated band structures and the corresponding DOS of pristine monolayer graphene (dots) and monolayer graphene with a (585) divacancy (solid). (b) Comparison between the experimental STS curves shown in fig. 4 for the divacancy B with the calculated DOS around the $E_{F}$.  }
\end{figure}

In summary, we present a comprehensive characterization of the geometrical and electronic structure of a (585) divacancy, a common point defect in irradiated graphene systems. This defect has a planar and non-hexagonal rearranged structure with no dangling bonds and thus it is expected to present a low chemical reactivity compared to other vacancy-type defects.  Diferential conductance spectra taken on divacancies generated at the surface of 4-5ML- graphene/SiC(000$\overline{1}$) reveal the existence of an electronic resonance centered at +150mV. The calculated band structure for the divacancy in a single layer of graphene fully supports the STS observations, proving the existence of a slightly dispersive electronic state of $\pi$ nature at this energy. Indeed, according to our DFT calculations, this type of divacancy does not show a magnetic character. The resonance associated to (585) divacancy at low energies is expected to limit the electron mobility \cite{Lherbier2011}, being thus an excellent candidate to be used to functionalize graphene by ion irradiation methods.

\begin{acknowledgments}
We thank Prof. J. M. Soler  and Dr. L. Magaud for fruitful discussions. This work was supported by Spain’s MICINN under grants No. MAT2010-14902, No. CSD2010-00024 and No. CSD2007-00050 and by Comunidad de Madrid under grant No. S2009/MAT-1467. M.M.U, I.B, F.H, P.M, J.Y.V and J.M.G.R also acknowledge the PHC Picasso program for financial support (project n 22885NH). M.M.U. acknowledges financial support from MEC under FPU Grant No. AP-2004-1896. I.B was supported by a Ram\'on y Cajal project of the Spanish MEC. F.H. held a doctoral support from the R\'egion Rh\^one-Alpes.
\end{acknowledgments}


\begin{thebibliography}{49}

\bibitem{Novoselov2005} K. S. Novoselov \emph{et al.}, Nature {\bf438}, 197 (2005).
\bibitem{Novoselov2004} K. S. Novoselov \emph{et al.}, Science {\bf306}, 666 (2004).
\bibitem{Adam2003} S. Adam, E. H. Hwang, V. M. Galitski, and S. Das Sarma, Proc.  Natl. Acad. Sci. USA {\bf104}, 18392 (2007).
\bibitem{Tan2007} Y. W. Tan \emph{et al.}, Phys. Rev. Lett. {\bf99}, 246803 (2007).
\bibitem{Katsnelson2008} M. I. Katsnelson and A. K. Geim, Phil. Trans. R. Soc.A {\bf366}, 195 (2008).
\bibitem{Titov2010} M. Titov \emph{et al.}, Phys. Rev. Lett. {\bf104}, 076802 (2010).
\bibitem{Ni2010} Z. H. Ni \emph{et al.}, Nano Letters {\bf10}, 3868 (2010).
\bibitem{Lusk2008} M. T. Lusk and L. D. Carr, Phys. Rev. Lett. {\bf100}, 175503 (2008).
\bibitem{Banhart2011} F. Banhart, J. Kotakoski, and A. V. Krasheninnikov, ACS Nano {\bf5}, 26 (2011).
\bibitem{Meyer2008} J. C. Meyer \emph{et al.}, Nano Letters {\bf8}, 3582 (2008).
\bibitem{Kelly1996} K. F. Kelly \emph{et al.}, Science {\bf273}, 1371 (1996).
\bibitem{Ruffieux2000} P. Ruffieux \emph{et al.}, Phys. Rev. Lett. {\bf84}, 4910 (2000).
\bibitem{Ugeda2010} M. M. Ugeda, I. Brihuega, F. Guinea, and J. M. G\'omez-Rodr\'iguez, Phys. Rev. Lett. {\bf104}, 096804 (2010).
\bibitem{Ugeda2011} M. M. Ugeda \emph{et al.}, Phys. Rev. Lett. {\bf107}, 116803 (2011).
\bibitem{Cockayne2011} E. Cockayne \emph{et al.}, Phys. Rev. B {\bf83}, 195425 (2011).
\bibitem{Esquinazi2003} P. Esquinazi \emph{et al.}, Phys. Rev. Lett. {\bf91}, 227201 (2003).
\bibitem{Ramos2010} M. A. Ramos \emph{et al.}, Phys. Rev. B {\bf81}, 214404 (2010).
\bibitem{Peng2008} X. Y. Peng and R. Ahuja, Nano Letters {\bf8}, 4464 (2008).
\bibitem{Lehtinen2004} P. O. Lehtinen \emph{et al.}, Phys. Rev. Lett. {\bf93}, 187202 (2004).
\bibitem{Vozmediano2005} M. A. H. Vozmediano, M. P. Lopez-Sancho, T. Stauber, and F. Guinea, Phys. Rev. B {\bf72}, 155121 (2005).
\bibitem{Pereira2006} V. M. Pereira \emph{et al.}, Phys. Rev. Lett. {\bf96}, 036801 (2006).
\bibitem{Amara2007} H. Amara \emph{et al.}, Phys. Rev. B {\bf76}, 115423 (2007).
\bibitem{Palacios2008} J. J. Palacios, J. Fern\'andez-Rossier, and L. Brey, Phys. Rev. B {\bf77}, 195428 (2008).
\bibitem{Yazyev2008} O. V. Yazyev, Phys. Rev. Lett. {\bf101}, 037203 (2008).
\bibitem{Zhang2011} X. Zhang \emph{et al.}, Carbon {\bf49}, 3615 (2011).
\bibitem{Krasheninnikov2001} A. V. Krasheninnikov \emph{et al.}, Phys. Rev. B {\bf63}, 245405 (2001).
\bibitem{Lehtinen2010} O. Lehtinen \emph{et al.}, Phys. Rev. B {\bf81}, 153401 (2010).
\bibitem{Lherbier2011}A. Lherbier \emph{et al.}, Phys. Rev. Lett. {\bf106}, 046803 (2011).
\bibitem{Gomez-Navarro2005}C. G\'omez-Navarro \emph{et al.}, Nature Mater. {\bf4}, 534 (2005).
\bibitem{Ugeda2011b} M. M. Ugeda, PhD, thesis, Universidad Aut\'onoma de Madrid, 2011.
\bibitem{Horcas2007} I. Horcas \emph{et al.}, Rev. Sci. Instrum. {\bf78}, 013705 (2007).
\bibitem{Kohn1965} W. Kohn and L. J. Sham, Phys. Rev. {\bf140}, A1133 (1965).
\bibitem{Ordejon1996} P. Ordej\'on, E. Artacho, J. D. Gale, and J. M. Soler, Phys. Rev. B {\bf53}, R10411 (1996).
\bibitem{Soler2002} J. M. Soler \emph{et al.}, J. Phys.- Cond. Matt. {\bf14}, 2745 (2002).
\bibitem{Sankey1989} O. F. Sankey and D. J. Niklewski, Phys. Rev. B {\bf40}, 3979 (1989).
\bibitem{Dion2004} M. Dion \emph{et al.}, Phys. Rev. Lett. {\bf92}, 246401 (2004).
\bibitem{Kong2009} L. Z. Kong, G. Rom\'an-P\'erez, J. M. Soler, and D. C. Langreth, Phys. Rev. Lett. {\bf103}, 096103 (2009).
\bibitem{Mallet2007} P. Mallet \emph{et al.}, Phys. Rev. B {\bf76}, 041403 (R) (2007).
\bibitem{Brihuega2008} I. Brihuega \emph{et al.}, Phys. Rev. Lett. {\bf101}, 206802 (2008).
\bibitem{Varchon2008} F. Varchon, P. Mallet, L. Magaud, and J. Y. Veuillen, Phys. Rev. B {\bf77}, 165415 (2008).
\bibitem{Hass2008} J. Hass \emph{et al.}, Phys. Rev. Lett. {\bf100}, 125504 (2008).
\bibitem{Li2010} G. H. Li \emph{et al.}, Nature Phys. {\bf6}, 44 (2010).
\bibitem{x} G. T. de Laissardiere, D. Mayou, and L. Magaud, Nano Letters {\bf10}, 804 (2010).
\bibitem{Sprinkle2009} M. Sprinkle \emph{et al.}, Phys. Rev. Lett. {\bf103}, 226803 (2009).
\bibitem{Pong2005} W. T. Pong and C. Durkan, J. Phys. D {\bf38}, R329 (2005).
\bibitem{Tersoff1983} J. Tersoff and D. R. Hamann, Phys. Rev. Lett. {\bf50}, 1998 (1983).
\bibitem{Lehtinen2011} O. Lehtinen, J. Kotakoski, A. V. Krasheninnikov, and J. Keinonen, Nanotechnology {\bf22}, 175306 (2011).


\end{thebibliography}
\end{document}